\documentclass[fleqn,usenatbib]{mnras}
\usepackage{newtxtext}
\usepackage{mathptmx}
\usepackage{mathtools}
\usepackage{txfonts}
\usepackage[T1]{fontenc}
\usepackage{ae,aecompl}
\usepackage{xcolor}

\DeclareRobustCommand{\VAN}[3]{#2}
\let\VANthebibliography\thebibliography
\def\thebibliography{\DeclareRobustCommand{\VAN}[3]{##3}\VANthebibliography}

\usepackage{graphicx}	
\usepackage{amsmath}	
\usepackage{amssymb}
\usepackage{bm}
\usepackage{pdflscape}

\title{Gamma-ray emission from primordial black hole-neutron star interaction}

\author[O. del Barco]{Oscar del Barco\thanks{E-mail: obn@um.es}
\\
Laboratorio de \'{O}ptica, Instituto Universitario de Investigaci\'{o}n en \'{O}ptica y
Nanof\'{i}sica, Universidad de Murcia, Campus de Espinardo, E-30100, Murcia, Spain\\}

\pubyear{2022}

\begin{document}
\label{firstpage}
\pagerange{\pageref{firstpage}--\pageref{lastpage}}
\maketitle

\begin{abstract}
The interaction of an asteroid-mass primordial black hole
(PBH) with a slowly-rotating neutron star (NS)
can lead to detectable gamma-ray emission via modern
observatories like Fermi-LAT or e-ASTROGRAM.
Depending on the specific PBH relativistic orbit in the
NS Schwarschild spacetime and the relative orientation
of this binary system with respect to Earth, the PBH
Hawking radiation will show a characteristic temperature
profile over time. Essentially, a moderate heating behaviour
(or even a progressive and constant cooling phase) is found
for the majority of the event, followed by a sudden and
dramatic cool-down at the end of the burst.
Our theoretical model might provide a means of identification
of such hypothetical PBH-NS interactions, based on the distinctive
temperature evolution of thermal-like gamma ray bursts (GRBs)
described in this article.
\end{abstract}


\section{Introduction}

Primordial black holes constitute an hypothetical type
of black holes (BHs) formed in the early Universe
through gravitational collapse of extremely dense regions
\citep{Zeldovich1967, Hawking1971, Carr1974}.
The study of such ultracompact objects of primitive
origin is a subject of continuous research
since it is thought that part or the totality of the dark matter
(DM) might be formed by PBHs of different masses
\citep{Inomata2017, Bartolo2019,
Belotsky2019, deLuca2021}.
In this regard, a significant part of the dark
matter in the Universe range in the
$10^{13} - 10^{14}$ kg, $10^{17} - 10^{21}$ kg
and $10 - 10^{3} M_{\sun}$ window,
not excluding lighter PBHs with less
contribution to the total amount of DM
\citep{Carr2020, Carr2021, Carr2022}.
It is also evident that these PBH masses are
significantly lower than the already evidenced
supermassive black holes \citep{Akiyama2019, Akiyama2022}.

Specifically, PBHs within the $10^{13} - 10^{14}$ kg
interval are constrained by observations of Galactic center
positrons and gamma-rays \citep{Laha2019, Laha2020, Dasgupta2020},
while new limitations on spinning and non-spinning PBHs
have been reported \citep{Saha2022}.
Larger PBHs in the solar-mass range have recently been associated
with gravitational wave signals \citep{Kohri2021} while PBHs smaller
than $10^{12}$ kg would have presently
evaporated due to its Hawking emission \citep{Ackermann2018}.
Current observations reported plausible PBH
detection via their Hawking radiation in the mass range
$8 \times 10^{11} - 10^{13}$ kg \citep{Wang2021}.
Moreover, ultra-light PBHs with masses less than $10^{6}$ kg
(which would have evaporated before big-bang nucleosynthesis)
might have been predominant in the early Universe
\citep{Papanikolaou2021}.

Among all of these different mass windows, the so-called
asteroid-mass PBHs (AMPBHs) have drawn increasing attention.
Some authors set the mass limit of these light BHs
within the mass interval $7 \times 10^{13} - 8 \times 10^{18}$ kg
\citep{Montero-Camacho2019} whereas other researchers
suggest PBHs with masses less than $2 \times 10^{19}$ kg
as a reasonable definition for AMPBHs \citep{Miller2022}.
In any case, these asteroid-mass PBHs
have lifetimes ranging from hundreds
to several million times the age of the Universe and
constitute a serious candidate for the cosmological dark matter
\citep{Coogan2021}. New constraints on this asteroid-mass window
have been lately reported \citep{Smyth2020, Miller2022},
being a matter of subsequent
research via future gamma ray telescopes
such as the upcoming AMEGO \citep{Ray2021}.

Recently, a plausible PBH origin for thermal GRBs
was stated, where an atomic-sized PBH described
a radial fall onto a massive black hole
\citep{Barco2021, Barco2022}. As a result, the
numerical calculations for the PBH
Hawking emission were consistent with thermal-like
gamma ray bursts (GRBs) in the MeV domain.
However, this study considered a quite restrictive
interaction between both black holes, that is,
a plunging orbit where the PBH is being captured
by its massive companion. No other possible relativistic
orbits were analyzed in this paper and a detailed discussion
of the rate of occurrence of such events was not provided.

In the present article, the above-mentioned astrophysical
scenario and the related theoretical model
are both updated to take into account a more general
PBH interaction event. Indeed,
an asteroid-mass PBH experiences a close approach to
a slowly-rotating neutron star
\citep{Yazadjiev2016, Motahar2017, Boumaza2021}
well approximated by a Schwarschild spacetime.
Two different relativistic orbits are studied in our
model, that is, a plunge case (where the PBH orbit
intercepts the NS surface) and a scattering event
around the NS, both with characteristic and recognizable
Hawking emission profiles. Moreover, a complete
analysis of the rate of occurrence of both events
(following \citet{Capela2013} formalism) is
carried out.

The paper is organized as follows. In section 2 we
describe our astrophysical scenario and develop
the theoretical framework to calculate the
PBH relativistic orbits and its fluence spectrum
$\nu S_{\nu}$ during the whole event.
A detailed numerical study concerning the PBH
plunge and scattering orbits and its associated
Hawking temperature evolution are performed
in section 3. A moderate
heating behaviour (or a progressive and constant
cooling, depending on the PBH orbit) is found for
the majority of the event, followed by a sudden and
dramatic cool-down at the end of the burst.
The gamma-ray emission due to the PBH Hawking
radiation might be detected via modern
observatories like Fermi-LAT or e-ASTROGRAM.
For completeness, the rate of occurrence
of such PBH-NS interactions is fully analyzed
in section 4. Finally, we summarize our
results in section 5.

\section{Theoretical basis}

\begin{figure*}
\begin{center}
\includegraphics[width=.77\textwidth]{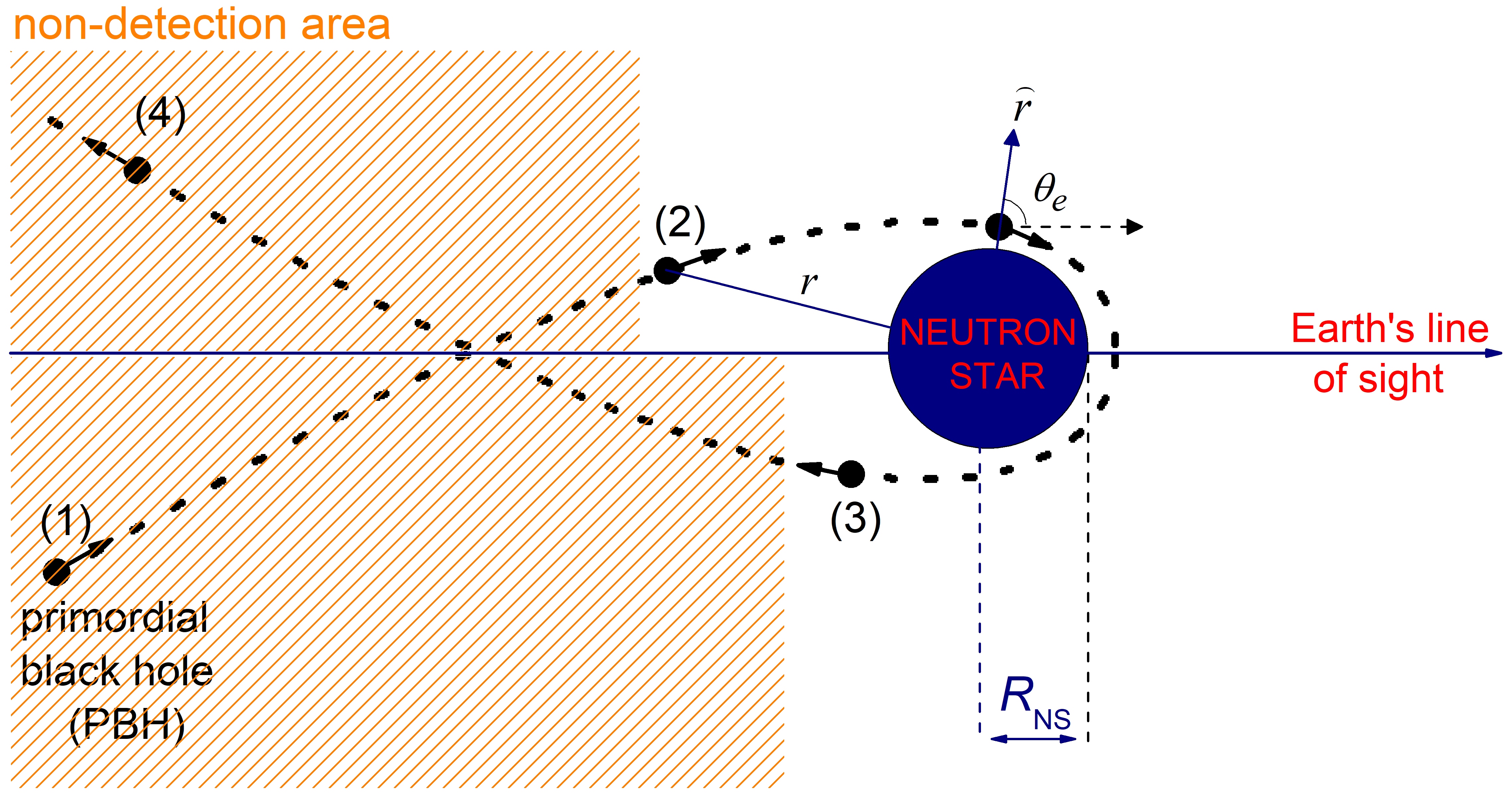}
\caption{Schematic representation of our
astrophysical scenario: an AMPBH
describes a relativistic orbit around an isolated
slowly-rotating NS. When the PBH is located
at the shadowed region (that is, the so-called
non-detection area), its Hawking radiation is too dim
to be detected via modern gamma-ray observatories.
As the PBH approaches the NS, its Hawking emission
is firstly Lorentz-boosted (please, see position (2))
and then gradually deboosted (position (3)) when it
completes its orbit around the compact object
(as in the case of a typical scattering orbit).
The parameter $\theta_e$ corresponds to the angle between
the radial coordinate $\hat{r}$ and the photon emission
direction (supposed along the Earth´s line of sight).}
\label{fig1}
\end{center}
\end{figure*}

Let us consider the astrophysical scenario depicted
in Fig.~\ref{fig1}. Here, an AMPBH of mass $M_{\rm P}$
describes a relativistic orbit around an isolated
slowly-rotating NS of radius $R_{\rm NS}$,
at distance $d_s$ from Earth.
It is also assumed that our NS does not accrete matter
from any stellar material, with such a low temperature
so that its proper thermal emission is practically
negligible \citep{Pearson2018, Pearson2019, Fantina2020}.

According to Fig.~\ref{fig1}, when the PBH is located
at position (1) (within the so-called non-detection area),
its Hawking emission lies below the typical sensibility
of modern observatories and, consequently, it cannot
be initially detected. Once the PBH approaches the NS
and its velocity component is pointing directly towards the Earth
(as in position (2)), its Hawking radiation is significantly
Lorentz-boosted and might be measured in the MeV range.
The parameter $\theta_e$ corresponds to the angle between
the radial coordinate $\hat{r}$ and the photon emission
direction (supposed along the Earth's line of sight, ELS).
As the PBH goes around the NS (in the case of a scattering
orbit), the velocity component is now
opposite to the ELS so a deboosting effect
occurs (please, see position (3) in Fig.~\ref{fig1}).
In this situation, the Hawking emission drops below
the gamma-ray observatories sensibility and enters
again the non-detection area.

Under these premises, the NS spacetime
can be properly modelled
by the Schwarschild line element
\begin{equation}\label{Schmetric}
ds^2 = - f(r) dt^2 + f(r)^{-1} dr^2 +
r^2 \left(d\theta^2 + \sin^2\theta \ d\phi^2\right),
\end{equation}
where $r$ indicates the radial coordinate
of the PBH (our emitter),
$f(r)=1-2M_{\rm NS}/r$ and
$M_{\rm NS}$ stands for the mass
of our NS. Trivially, for a distant observer
placed at $r_{\rm obs}$
(for example, an observatory on Earth), the
parameter $f(r_{\rm obs})\simeq 1$.

Throughout the article, the geometrized
system of units ($c = G = 1$) is adopted,
except for radiometric magnitudes where the
S.I. system of units is considered.
We also assume that the PBH motion
is confined at the equatorial plane
$\theta=\pi/2$ with conserved specific energy
$E_{s}$ and positive
specific angular momentum $L_{s}$.
Both parameters will determine the specific
PBH orbit, as will be briefly discussed.

On the other hand, the kinematic component
of the time dilation $\gamma$
can be expressed as \citep{Hartle2003}
\begin{equation}\label{gammal}
\gamma = \left(1-\beta^2\right)^{-1/2}=
\frac{E_{s}}{\gamma_{G}},
\end{equation}
where $\beta$ is the velocity of the PBH and
$\gamma_{G}=\sqrt{f(r)} / \sqrt{f(r_{\rm obs})}
\simeq \sqrt{f(r)}$. We can also write
the coordinate time $t$ (for a distant observer)
and the azimuthal
angle $\phi$ as a function of the PBH
radial coordinate $r$ via the following
differential equations \citep{Hartle2003}
\begin{equation}\label{drdt}
dt = \frac{E_{s}}{f(r)} \left[E_{s}^2-f(r)
\left(1+\frac{L_{s}^2}{r^2}\right)\right]^{-1/2} dr,
\end{equation}
and
\begin{equation}\label{drdphi}
d\phi = \frac{1}{r^2} \left[\frac{1}{b^2}-f(r)
\left(\frac{1}{L_{s}^2} +
\frac{1}{r^2}\right)\right]^{-1/2} dr,
\end{equation}
where $b=L_{s}/E_{s}$. After numerical resolution
of both expressions~(\ref{drdt}) and (\ref{drdphi}),
the event duration (from the point of view of an
observer on Earth) and the PBH specific orbit
will be respectively derived.

Furthermore, the PBH Hawking temperature
as measured by the emitter ($T_{\rm P}^{'}$)
and a distant observer ($T_{\rm P}$) are related
via the redshift (blueshift) factor $\alpha$ \citep{McMaken2022}
\begin{equation}\label{Tp}
T_{\rm P} = \alpha T_{\rm P}^{'} =
\gamma_{\rm G} \left(\gamma
\pm \cos\theta_{e} \sqrt{\gamma^2
-\left(1+\frac{L_{s}^2}{r^2}\right)} \mp
\frac{L_{s}}{r}\sin\theta_{e}\right)^{-1} T_{\rm P}^{'},
\end{equation}
where the upper (bottom) signs
stand for an infalling (outgoing) relativistic
orbit (such as the scattering case described
in Fig.~\ref{fig1}) and $T_{\rm P}^{'}$
can be well approximated by
\citep{Hawking1974, Hawking1975}
\begin{equation}\label{TPBH}
T_{\rm P}^{'} = \left(6.169
\times 10^{-8}\right)
\frac{M_{\sun}}{M_{\rm P}},
\end{equation}
where the temperature is expressed in Kelvins.
One notices that the PBH Hawking
temperature $T_{\rm P}$ depends on its
mass (that is, lighter PBHs would be
at higher temperature), the constants
of motion $E_{s}$ and $L_{s}$ (i.e., the specific
PBH orbit in the NS spacetime)
and the photon emission direction $\theta_{e}$
(or, equivalently, the position of the PBH
in its relativistic orbit).

Considering a slowly-rotating PBH,
it is thus appropriate to model the Hawking emission
as a blackbody radiator \citep{Murata2006, Barco2021},
so the fluence spectrum $\nu S_{\nu}$ for a static
observer on Earth can be expressed as (S.I. units)
\begin{equation}\label{nuSnu}
\nu S_{\nu} = \frac{2\pi h \nu^{4} \Omega}
{c^{2}} \left(\exp\left[\frac{h\nu}
{k T_{\rm P}}\right]-1\right)^{-1},
\end{equation}
and $\Omega = \pi r_{\rm D}^2 / d_{s}^2$
is the solid angle subtended by a
point source for a detector of radius $r_{\rm D}$
(which, in turn, is proportional to the
effective area $A_{\rm eff}$
of our gamma-ray telescope).
As discussed in the next section,
our astrophysical scenario predicts
a fluence spectrum $\nu S_{\nu}$
of finite duration lying within
the MeV range.

\section{Numerical results}

In this section, a complete numerical analysis
of the PBH relativistic orbits
and the associated Hawking emission (in connection
with thermal-like GRBs) is carried out.

Firstly, let us analyze the different PBH scenarios
allowed in our model within the theoretical
framework of the NS Schwarschild spacetime.
Without loss of generality, the mass
of our PBH will be $10^{12}$ kg throughout
the article (that is, in the range of the
aforementioned asteroid-mass interval \citep{Miller2022}).
As already stated, the specific PBH orbit depends
on the conserved magnitudes $E_{s}$ and $L_{s}$
via the effective potential $V_{\rm eff}$ \citep{Hartle2003}
\begin{equation}\label{Veff}
V_{\rm eff}(r)=\frac{1}{2} \left[f(r)
\left(1+\frac{L_{s}^2}{r^2}\right)-1\right].
\end{equation}

When $(L_{s} / M_{\rm NS}) > \sqrt{12}$,
the effective potential has one maximum
and one minimum: this will be the situation
considered in our model.
Moreover, the specific orbit depends on the
relationship between the parameter
$\epsilon=(E_{s}^2 -1)/2$ and $V_{\rm eff}$.
If $\epsilon>V_{\rm eff}^{(\rm max)}$, the PBH describes
a plunge orbit onto the NS (and, eventually,
it might be captured by the latter, as remarked in
section 4). When $0<\epsilon<V_{\rm eff}^{(\rm max)}$,
a scattering orbit occurs where the PBH comes in from
infinity, orbits the NS and moves out to infinity again
(as illustrated in Fig.~\ref{fig1}).

\begin{figure}
\begin{center}
\includegraphics[width=.49\textwidth]{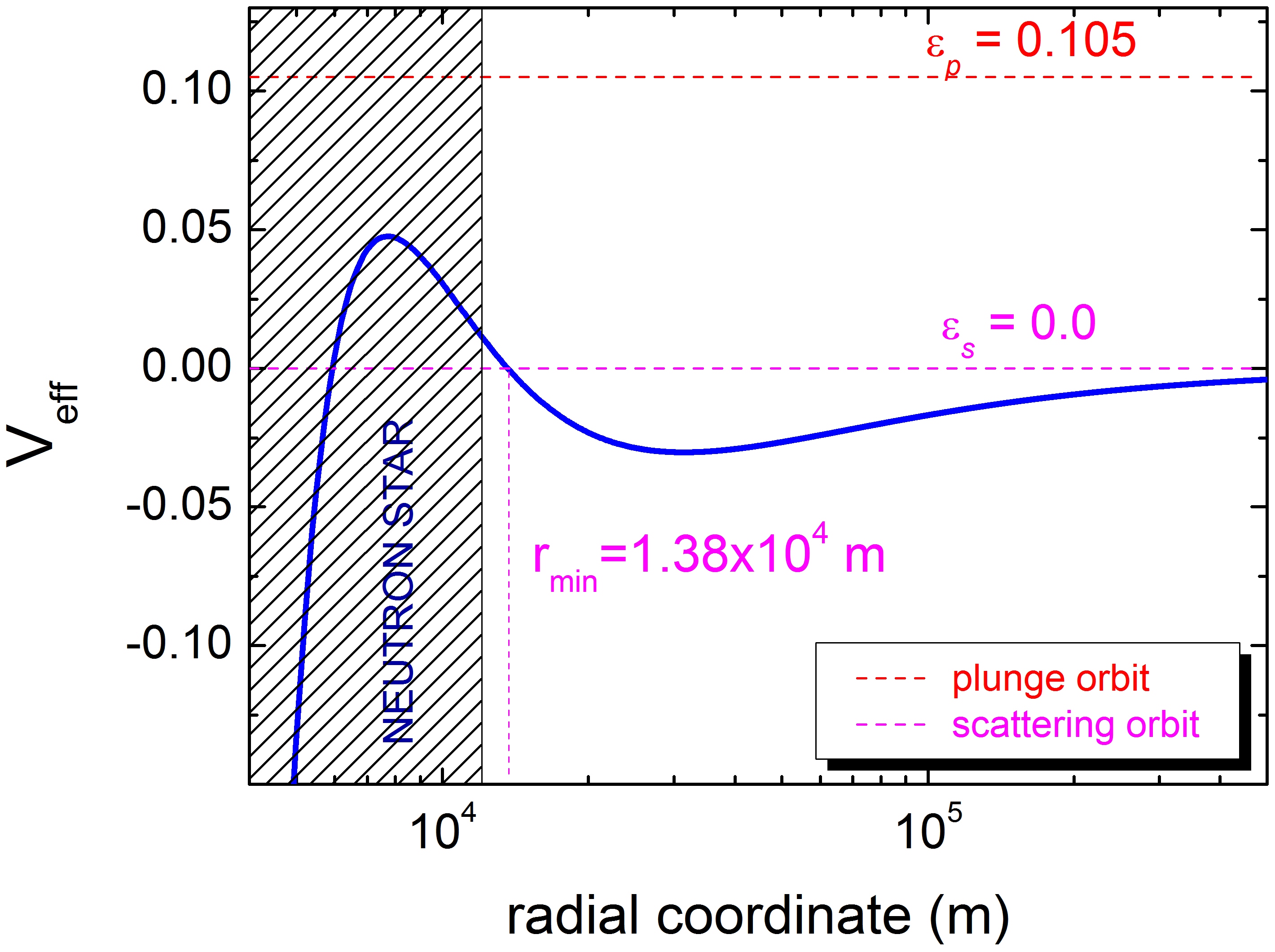}
\caption{The effective potential $V_{\rm eff}$
versus the radial coordinate $r$ (calculated via
equation~(\ref{Veff})) for a NS of mass
$M_{\rm NS}=1.4 M_{\sun}$ and radius
$R_{\rm NS} = 1.2 \times 10^4$ m
(considering that the slowly-rotating
NS spacetime is well-approximated
by the Schwarschild metric).
The selected value for the PBH specific
angular momentum was $L_{s}=9 \times 10^3$ m,
so $V_{\rm eff}$ presents one maximum
and one minimum. Two different values
of the parameter $\epsilon$
are shown, each representing
a characteristic PBH orbit:
a plunge event for $\epsilon_{p}=0.105$,
corresponding to a PBH specific energy $E_{s}=1.1$
and a velocity at infinity of 0.41c, and a scattering
orbit for $\epsilon_{s}=0.0$,
where the PBH exhibits a free fall from infinity.
For the scattering orbit, the minimum
approach of the PBH to the NS resulted to be $r_{\rm min}
= 1.38 \times 10^4$ m.}
\label{fig2}
\end{center}
\end{figure}

The effective potential $V_{\rm eff}$ is calculated
via equation~(\ref{Veff}) and represented
in Fig.~\ref{fig2} for a typical neutron star of
mass $M_{\rm NS}=1.4 M_{\sun}$ and radius
$R_{\rm NS} = 1.2 \times 10^4$ m.
The selected value for the PBH specific angular
momentum was $L_{s}=9 \times 10^3$ m.
With this choice, the NS effective
potential presents one maximum and one minimum, though
this prerequisite is not essential for our model
(as in the case when $(L_{s} / M_{\rm NS}) < \sqrt{12}$
where only plunge orbits are allowed).
Two different values of the parameter $\epsilon$
are shown in Fig.~\ref{fig2}, each representing
a characteristic PBH orbit:
a plunge event for $\epsilon_{p}=0.105$,
corresponding to a PBH specific energy $E_{s}=1.1$
and a velocity at infinity of 0.41c, and a scattering
orbit for $\epsilon_{s}=0.0$, where the PBH
exhibits a free fall from infinity.
In this case, the intercept of
$\epsilon$ with the effective potential
curve gives us the minimum approach of
the PBH to the NS (that is, $r_{\rm min}
= 1.38 \times 10^4$ m).

\begin{figure}
\begin{center}
\includegraphics[width=.49\textwidth]{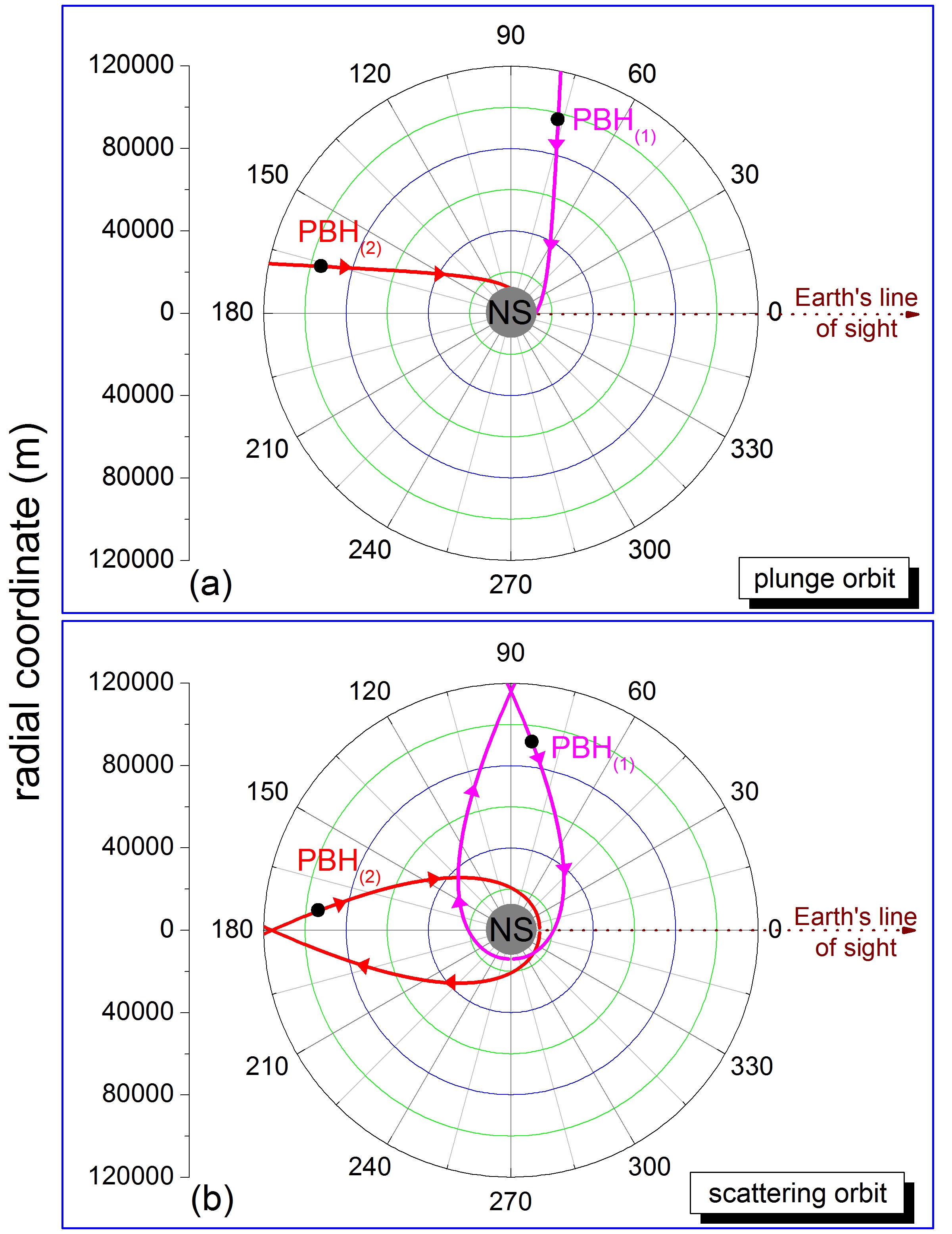}
\caption{Polar diagrams for the radial $r$
and azimuthal $\phi$ coordinates (last
one in degrees) obtained via numerical
resolution of equation~(\ref{drdphi}).
Two different AMPBH scenarios are considered
(same parameters as in Fig.~\ref{fig2}):
(a) plunge and (b) scattering orbits.
For each panel, two different relative
orientations of the PBH with respect to the ELS
are also shown (and denoted as $\textrm{PBH}_{(1)}$
and $\textrm{PBH}_{(2)}$, respectively).}
\label{fig3}
\end{center}
\end{figure}

In this context, the specific shape of each PBH orbit
(obtained via numerical resolution of
equation~(\ref{drdphi})) is represented
in Fig.~\ref{fig3}, where the radial $r$
and azimuthal $\phi$ coordinates
(last one in degrees) are plotted as polar
diagrams in the equatorial plane $\theta=\pi/2$.
Top panel depicts two identical
AMPBHs describing
different plunging orbits (in relation
to their relative orientation with
respect to the ELS,
and denoted as $\textrm{PBH}_{(1)}$
and $\textrm{PBH}_{(2)}$, respectively),
while bottom panel shows two scattering
events for the same astrophysical
scenario. In both cases, the same parameters
as in Fig.~\ref{fig2} have been selected.
In connection with Fig.~\ref{fig1}, the
infalling PBH approaching the NS from
infinity is not initially detected
(due to its faint Hawking emission).
Depending on its particular orbit,
the PBH Hawking radiation will be
rather different and recognizable,
as briefly discussed.

\begin{figure}
\begin{center}
\includegraphics[width=.48\textwidth]{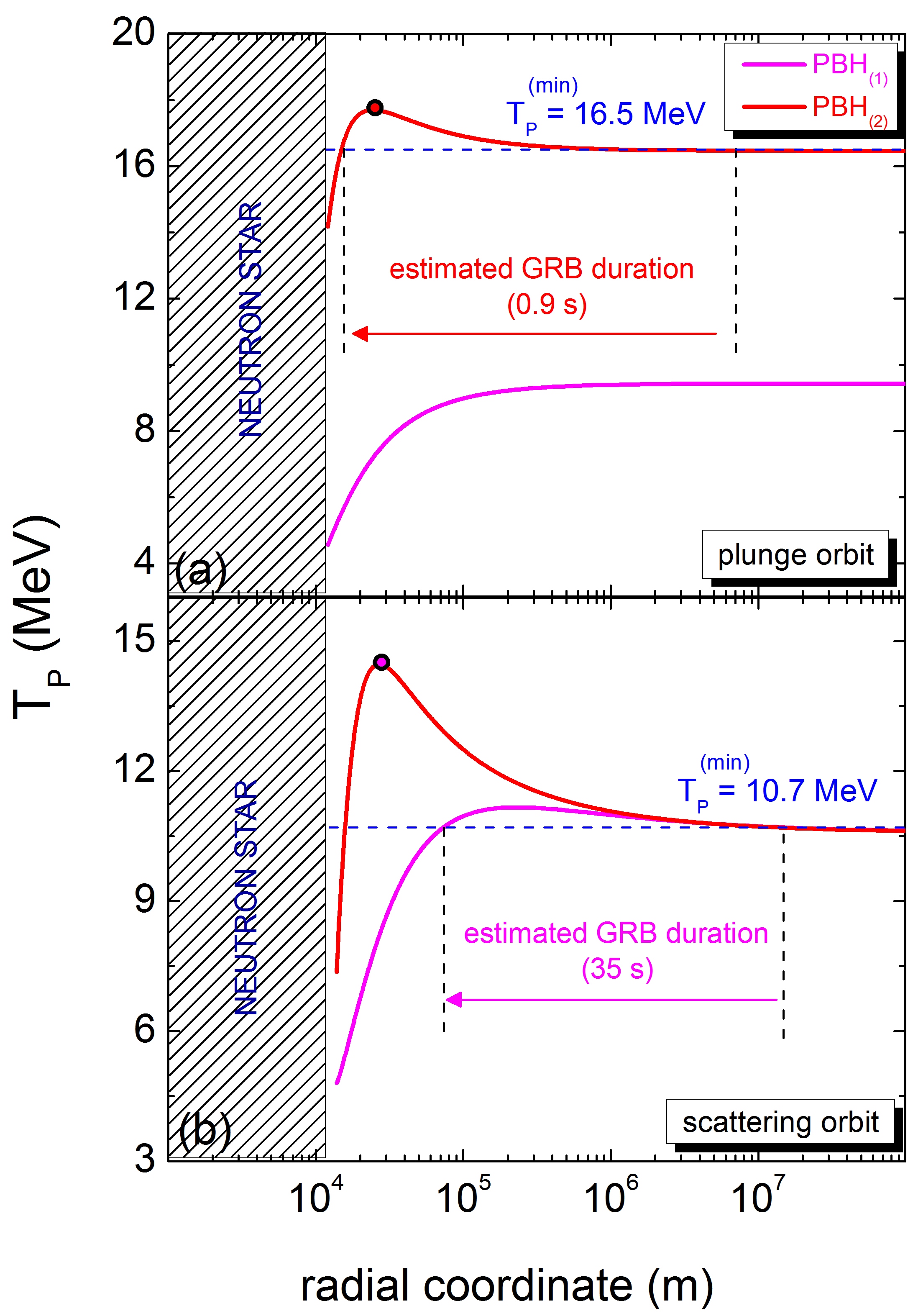}
\caption{PBH Hawking temperature $T_{\rm P}$ versus
the radial coordinate $r$ for each specific orbit,
calculated numerically via equation~(\ref{Tp})
(same parameters as in Fig.~\ref{fig3}).
The threshold value of the PBH Hawking temperature
$T_{\rm P}^{(\rm min)}$ detectable by modern
observatories on Earth is represented as dashed
horizontal lines in both panels: (a) plunge orbits,
where the $\textrm{PBH}_{(2)}$ event
might be measured as a thermal-like GRB of 0.9 s duration
and (b) scattering cases, with an estimated time
span of 35 s for $\textrm{PBH}_{(1)}$.}
\label{fig4}
\end{center}
\end{figure}

In Fig.~\ref{fig4}, the PBH Hawking temperature
$T_{\rm P}$ is illustrated as a function
of the radial coordinate $r$ for each particular
orbit in Fig.~\ref{fig3}. All our numerical
calculations have been performed via equation~(\ref{Tp}).
For both plunge and scattering events,
there exists a minimum value of the PBH Hawking
temperature $T_{\rm P}^{(\rm min)}$ (i.e., the dashed
horizontal lines in top and bottom panels in Fig.~\ref{fig4})
below which the fluence spectrum $\nu S_{\nu}$ is too dim
to be detected by modern observatories.
It should be also pointed out that this
temperature threshold strongly depends on the PBH mass,
its distance to the Earth (or, equivalently, the solid
angle subtended by such point source) and the sensitivity of the
gamma-ray telescopes. Once our PBH is close enough to
the NS, its Hawking temperature exceeds $T_{\rm P}^{(\rm min)}$
and its emission is likely to be measured: this is the beginning
of the thermal GRB of finite duration (and calculated
after numerical resolution of
differential equation~(\ref{drdt}),
as shortly presented).
This prompt gamma-ray emission is concluded when
$T_{\rm P}$ drops below the threshold temperature
$T_{\rm P}^{(\rm min)}$ again, a process which is highly
dependent on the PBH relativistic orbit.
Indeed, the $\textrm{PBH}_{(2)}$ orbit in
Fig.~\ref{fig4}(a) will produce a detectable
GRB of 0.9 s, while the
plunge orbit for $\textrm{PBH}_{(1)}$
will remain unnoticed for current
observatories.
On the other hand, both scattering orbits
in Fig.~\ref{fig4}(b) will be experienced as
thermal-like GRBs of specific duration (about 35
s for $\textrm{PBH}_{(1)}$).

It can be noticed that
the PBH Hawking temperature presents an initial
heating behaviour (due to the Lorentz-boosting of
$T_{\rm P}$ when the PBH velocity component is
directed towards the ELS), followed by a dramatic
cool-down, when the PBH velocity component
is opposite to the Earth’s line of sight
(and a Lorentz-deboosting process occurs).
This heating evolution is more pronounced
(and eventually reaching a well-defined maximum
temperature) for relativistic orbits where the
PBH velocity component is mainly maintained towards
us during practically the whole event
(please, see again the
$\textrm{PBH}_{(2)}$ cases in Fig.~\ref{fig4}).
In the opposite scenarios (as in the
$\textrm{PBH}_{(1)}$ events), there exists a
moderate heating process (such as the
scattering orbit in bottom panel) or even a
progressive and constant cooling,
as in the plunge case in Fig.~\ref{fig4}(a).
The latter situation is more consistent with
already reported cooling behaviour of thermal GRBs
\citep{Ryde1999, Ryde2004, Barco2021, Barco2022}.
It should be added that, as the PBH travels away
from the NS in the scattering case,
its Hawking temperature (as calculated via
equation~(\ref{Tp}), under the appropriate
choice of signs) continues its fall and never goes
above the threshold temperature $T_{\rm P}^{(\rm min)}$.
Provided the irrelevance of such results
to the detectability of scattering events,
these computational values have not
been included in Fig.~\ref{fig4}(b).

\begin{figure}
\begin{center}
\includegraphics[width=.48\textwidth]{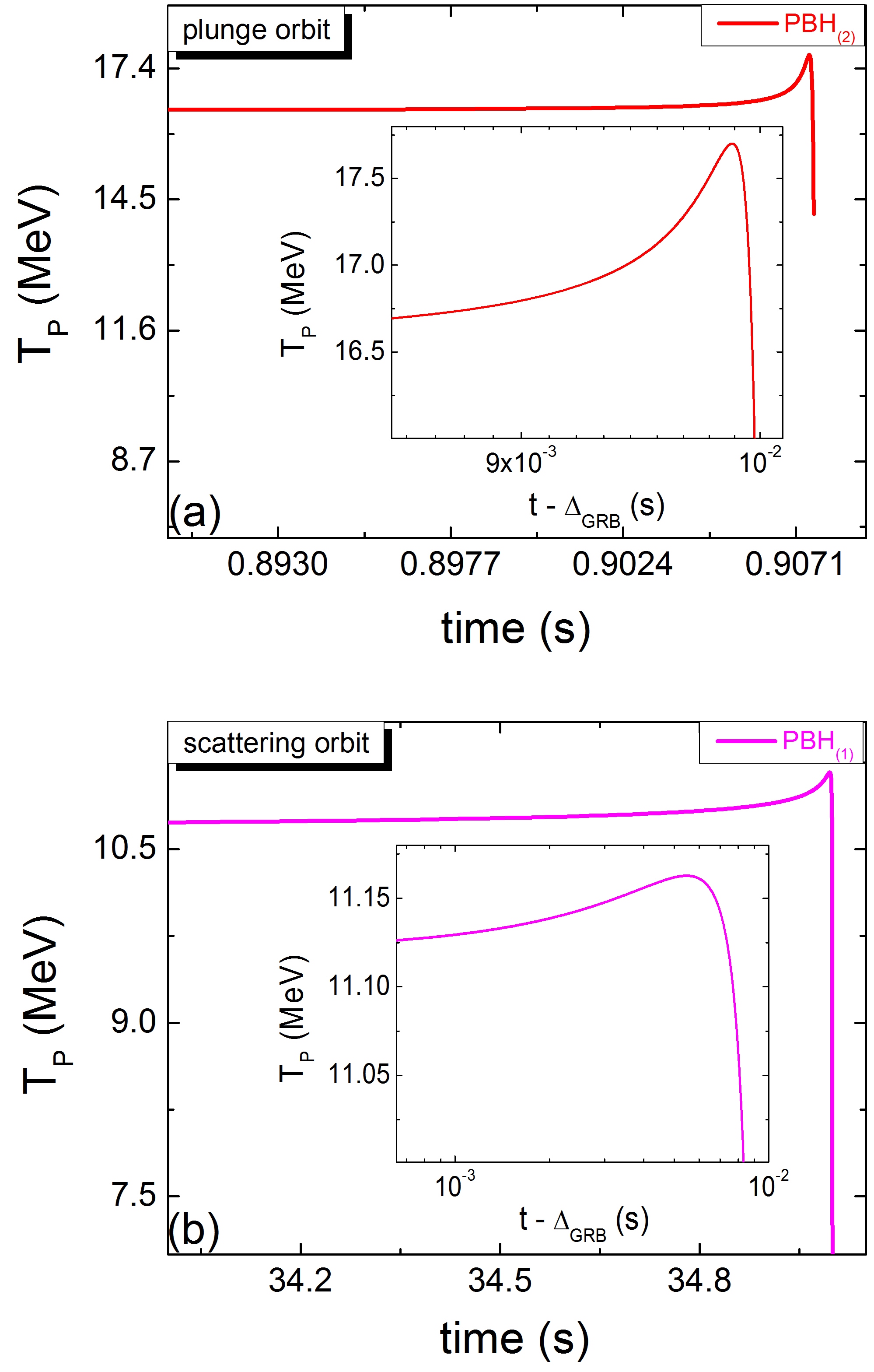}
\caption{PBH Hawking temperature $T_{\rm P}$ as a function
of the observer's coordinate time $t$, numerically
calculated via equation~(\ref{drdt})
(same parameters as in Fig.~\ref{fig4}).
For a better visualization of the rapid
cooling phase in both panels,
two insets have been included where
the horizontal axis describes the coordinate
time $t$, once the estimated GRB duration
$\Delta_{\rm GRB}$ is subtracted.}
\label{fig5}
\end{center}
\end{figure}

The corresponding temporal evolution of the
PBH Hawking temperature $T_{\rm P}$ is shown in
Fig.~\ref{fig5} (via numerical resolution
of equation~(\ref{drdt}) and same parameters
as in Fig.~\ref{fig4}). For the
$\textrm{PBH}_{(2)}$ plunge case
(upper panel) and the $\textrm{PBH}_{(1)}$ scattering
event (bottom panel), a moderate heating behaviour is obtained
followed by a dramatic temperature
drop of millisecond duration,
as depicted in the insets (where the abscissa axis
represents the coordinate time $t$,
once the estimated GRB duration $\Delta_{\rm GRB}$
is subtracted). This is probably the new signature
reported in this article, where the cooling phase
is much too steep compared to the heating one.

\begin{figure}
\begin{center}
\includegraphics[width=.47\textwidth]{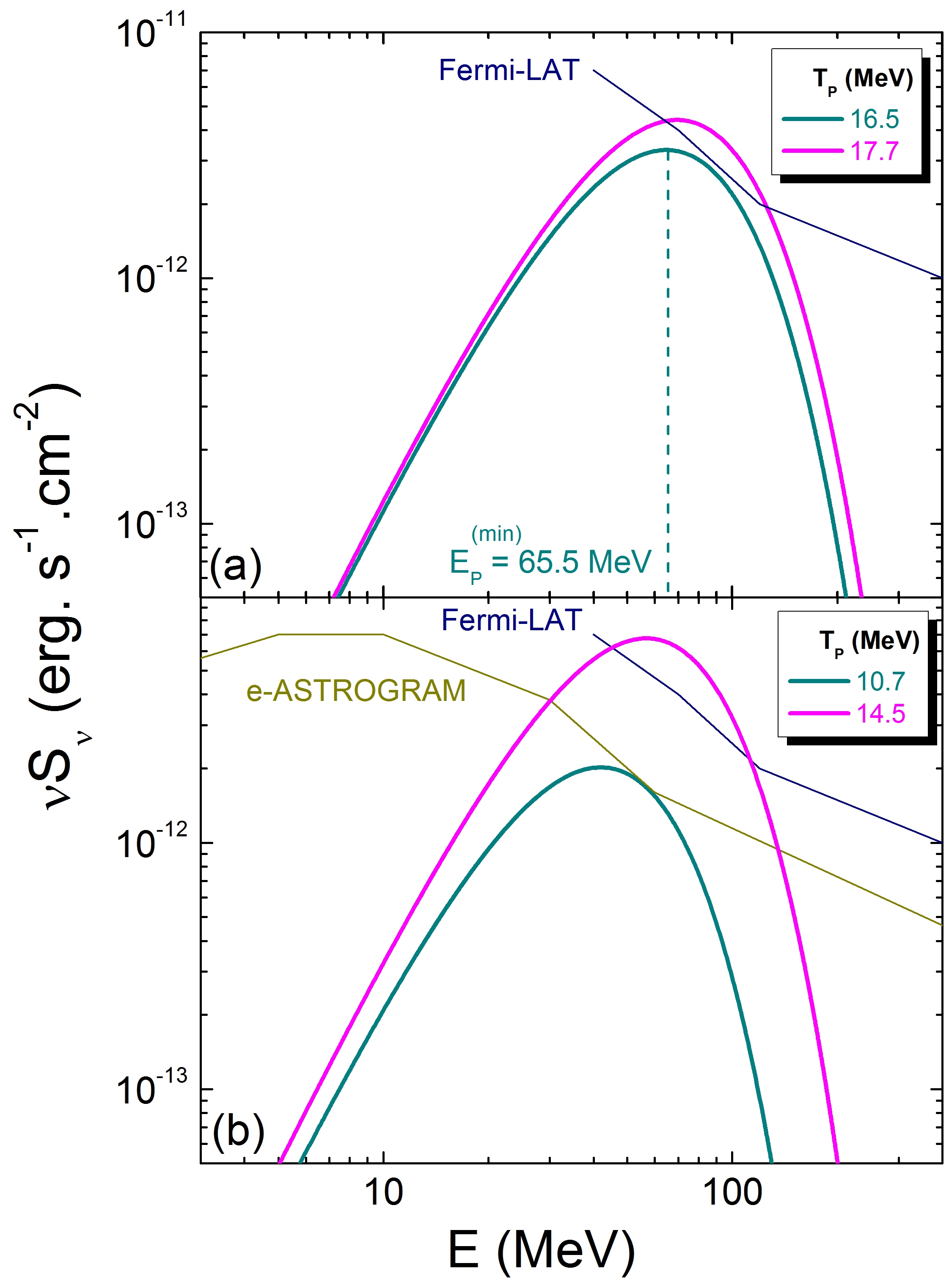}
\caption{Fluence spectrum $\nu S_{\nu}$
for the minimum and maximum PBH Hawking
temperatures (same parameters
as in Figs.~\ref{fig3}
and \ref{fig4}) for the $\textrm{PBH}_{(2)}$
events. Our numerical results were calculated via
equation~(\ref{nuSnu}) for (a) plunge
and (b) scattering orbits.
It can be observed a typical
Planckian-like distribution for $\nu S_{\nu}$,
in accordance with the PBH thermal emission.
As an indicative manner, the sensitivities
of the e-ASTROGRAM and Fermi-LAT telescopes
are shown.}
\label{fig6}
\end{center}
\end{figure}

In relation to the fluence spectrum
$\nu S_{\nu}$ plausibly measurable by modern
gamma-ray stations, Fig.~\ref{fig6}
shows this magnitude for the
$\textrm{PBH}_{(2)}$ events and calculated
via equation~(\ref{nuSnu})
(please, see again Figs.~\ref{fig3}
and \ref{fig4}). It can be observed a typical
Planckian-like distribution for $\nu S_{\nu}$,
in accordance with the PBH thermal emission.
For the plunge orbit case (top panel), the fluence
spectrum corresponding to a minimum Hawking temperature
of 16.5 MeV practically reaches the sensitivity of the Fermi-LAT
observatory \citep{deAngelis2018}. For this
calculation, an estimated distance to Earth of 4.2 Gparsec
and an effective area
$A_{\rm eff}=10^4$ $\textrm{cm}^{2}$ for the
Fermi-LAT \citep{Coogan2021} have been considered.
Furthermore, the PBH fluence spectrum $\nu S_{\nu}$
for the maximum Hawking temperature 17.7 MeV (i.e., the red dot
shown in Fig.~\ref{fig4}(a)) is also represented,
where now $\nu S_{\nu}$ is well above
the Fermi-LAT sensitivity.

In Fig.~\ref{fig6}(b), the scattering orbit event is
depicted for a PBH-NS system at 2.3 Gparsec from Earth.
One notices that the minimum fluence spectrum (associated with
the threshold temperature
$T_{\rm P}^{(\rm min)}=10.7$ MeV, please see again
Fig.~\ref{fig4}(b)) is of the order
of the e-ASTROGRAM observatory sensibility \citep{deAngelis2018},
with an estimated effective area
$A_{\rm eff}=2 \times 10^3$ $\textrm{cm}^{2}$
in the MeV domain \citep{Coogan2021}.
The maximum value for $\nu S_{\nu}$ (corresponding
to a PBH temperature of 14.5 MeV,
as illustrated in Fig.~\ref{fig4}(b)
with a red dot) peaks above the Fermi-LAT sensitivity,
being detectable at the final stage of the burst
(just before the cooling phase), via such modern observatories.

At the same time, it should be recalled
that our slowly-rotating NS does not possess
an accretion disk
\citep{Pearson2018, Pearson2019, Fantina2020},
unlike other younger and more active NSs.
The presence of such structure
in our binary astrophysical
scenario might modify our previous
fluence calculations, due to the thermal
emission of this accretion disk.
Additionally, as previously commented,
our isolated NS has
cooled down sufficiently so that
its proper thermal emission
is negligible.

To complete this section, it is worth
asking how frequent are such PBH-NS
interactions, in order to be
detectable by gamma-ray
observatories. We will try to
give a satisfactory response on
this topic in the next section.

\section{Rate of occurrence analysis}

Let us now examine the rate of
occurrence $\mathcal{F}$ of such PBH-NS
scenarios, including the already studied
plunge and scattering cases.
Such astrophysical events have been earlier
reported in the literature where (in a more
restrictive situation), a PBH is captured by a neutron star
\citep{Capela2013, Pani2014}.
In this way, it is a reasonable assumption that
the PBH-NS events leading to the above-mentioned
GRBs should be of extragalactic nature
(otherwise, nearby galaxies would be a remarkable
source of brighter gamma ray transients,
a possibility that must be discarded).

Following the formalism by \citet{Kouvaris2008} and
\citet{Capela2013}, the parameter $\mathcal{F}$
is derived by integrating a Maxwellian dark matter
distribution, with velocity dispersion $\bar{v}$,
over the $E_{s}$ and $L_{s}$ space with
well-defined integration limits.
Within this theoretical framework,
a PBH is captured and becomes gravitationally bound when
it loses its initial energy due to the
accretion of star's material.
Nevertheless, we would not adopt such
limiting situation: it would be enough for us
to consider that the PBH orbit
intercepts the NS surface (as in the case of a plunge
event) or that the PBH passes close enough to
the NS (for a scattering scenario).

Accordingly, the parameter $\mathcal{F}$ for a
NS of radius $R_{\rm NS}$
can be written as \citep{Capela2013}
\begin{equation}\label{cr1}
\mathcal{F} = \sqrt{6\pi} \frac{\Omega_{\rm PBH}}{\Omega_{\rm DM}}
\frac{\rho_{\rm DM}}{M_{\rm P}}
\frac{R_{\rm NS} R_{s}}{\bar{v} f(R_{\rm NS})},
\end{equation}
where $\Omega_{\rm PBH}/\Omega_{\rm DM}$ is the fraction
of PBHs to dark matter, $\rho_{\rm DM}$ the local DM density and
$R_{s}= 2M_{\rm NS}$ the Schwarschild radius
of the NS.
Assuming that the currently estimated DM density
ranges from low density regions
(where $\rho_{\rm DM} \simeq 0.5$ $\mathrm{GeV.cm^{-3}}$
at the edge of a galaxy) to high values around
840 $\mathrm{GeV.cm^{-3}}$ at the center of a standard
galaxy \citep{Read2014, Sofue2020, Chang2021},
a velocity dispersion
$\bar{v}=7 \times 10^3$ $\mathrm{m.s^{-1}}$
and typical NS parameters
(i.e., $R_{\rm NS} = 1.2 \times 10^4$ m
and $M_{\rm NS} = 1.4 M_{\sun}$),
the parameter $\mathcal{F}$ for an AMPBH
of $10^{12}$ kg describing a plunge orbit
(and calculated via equation~(\ref{cr1}))
should be of $10^{-3} \mathrm{yr^{-1}}$.
Considering a fraction of dark matter in the form
of asteroid-mass PBHs as
$\Omega_{\rm PBH}/\Omega_{\rm DM} \simeq 10^{-7}$
\citep{Bartolo2019, deLuca2021}, the rate of occurrence
for a single PBH plunge event corresponds to
$\mathcal{F} \simeq 10^{-10} \mathrm{yr^{-1}}$.

Taking into account an estimated number of
$10^{8}$ NSs in a galaxy similar to
the Milky Way \citep{Sartore2010, Cordes2016}
and a specific area of the Universe
with $10^{3}$ galaxies at distances greater
than 1 Gparsec from Earth (among $10^{11}$ galaxies
in the observable Universe, as reported by
\citet{Conselice2006, Conselice2016, Lauer2021}),
the number of such AMPBH plunge events
that might occur at about $10^{3}$
distant galaxies should be roughly 12 events per
year (assuming an mean value of
$\rho_{\rm DM} \simeq 5$ $\mathrm{GeV.cm^{-3}}$
in our calculations).
For the scattering case,
the rate of occurrence
$\mathcal{F}$ is of the order of 19 events per year
(where now the NS radius has
been expanded to include such orbits, concretely
$R_{\rm NS} = 2.4 \times 10^4$ m).

It is also worth recalling that many
of these PBH-NS interactions would not give
rise to observable GRBs, mainly due to
the small range of successful orientations
of our binary system.
Moreover, the differences
in PBH concentration and DM densities in
the Universe (excluding young NSs in the galaxies)
could be another limiting factor.
Despite the difficulty of finding PBH-NS interactions
leading to detectable GRBs (and consistent with our
theoretical predictions), those astrophysical scenarios
should not be discarded, as already discussed in this section.

\section{Conclusions}

Summarizing, an astrophysical scenario
where an asteroid-mass PBH undergoes
a scattering or a plunge orbit
towards a slowly-rotating
NS can lead to detectable gamma-ray
emission, via modern observatories
like Fermi-LAT or e-ASTROGRAM.
As widely studied in the previous sections,
the PBH Hawking radiation
shows a well-defined
and characteristic temperature profile over time,
depending on the specific
PBH event in the NS Schwarschild spacetime.
Related to this, the PBH stable circular
orbit (that is, when the parameter
$\epsilon$ equals the minimum
value of $V_{\rm eff}$, please see
again Fig.~\ref{fig2}) has not
been discussed in our model.
In this hypothetical scenario,
a periodic thermal GRB of short
duration should be measured, in addition
to the earlier described signatures
in section 3.

No afterglows are expected for the
scattering case, where our PBH goes
around the NS and moves away again to infinity
(that is, a dark GRB would occur
\citep{Fynbo2001, Jakobsson2004, Melandri2012}).
Nonetheless, subsequent thermal
emission at longer wavelengths
(i.e., afterglows within the X-ray or
ultraviolet domain) might happen
for a plunge event due to
the accretion of heated stelar
material, once the PBH is trapped
by the NS. This accretion mechanism
should be more efficient for more massive
incoming primordial black holes
(such as planetary-mass PBHs)
where their event horizon is
much greater than the size
of an atom: let us recall
that the size of an AMPBH
(like the one considered in our work)
is of the order of an atomic nucleus,
roughly $10^{-15}$ m.

As commented in section 4,
the rate of occurrence
of these PBH-NS interactions
(within a specific region of the Universe,
covering $10^{3}$ galaxies
at distances greater than
1 Gparsec from Earth)
should be of about
12 or 19 events per year, depending on
the plunge or scattering case, respectively.
However, all of these events
would not produce detectable
GRBs, mainly due to the difficulties
in the correct alignment of the binary system
(that is, only a small range
of orientations in our theoretical model
could successfully generate such measurable signatures).
Moreover, as previously commented,
the interaction events resulting in detectable
thermal-like GRBs would occur at cosmological
distances (that is, when such
asteroid-mass PBHs were more
abundant than presently),
constraining thus the number of PBH-NS
plausible scenarios.
It can also be expected that future
gamma-ray observatories with improved
sensitivities might be able to detect
such GRBs of primordial black hole origin.

On the other hand, similar results
to those detailed in this article
are expected if we consider a
stellar-mass black hole as
the massive central object,
instead of a NS. In this context,
the asteroid-mass PBH could get
through the BH photon sphere and,
as a consequence, the cool-down process
might be more pronounced and accelerated
\citep{Barco2022}.

In conclusion, our theoretical model
might provide a way of identifying such
feasible PBH-NS interactions, based on
the specific temperature profile of the
before-mentioned thermal-like GRBs.

\section*{Acknowledgements}

The author gratefully acknowledges the anonymous reviewer
for valuable assistance throughout the refereing process.
O. del Barco warmly acknowledges F. J. \'{A}vila
for helpful discussions on neutron star evolution
and thanks research support from
University of Zaragoza- Fundaci\'{o}n Ibercaja
(project number JIUZ-2021-CIE-01) and
Fundaci\'{o}n S\'{e}neca -
Agencia de Ciencia y Tecnolog\'{i}a  de
la Regi\'{o}n de Murcia (19897/GERM/15).

\section*{Data Availability}

Provided the theoretical nature of this paper,
all data and numerical results generated or analysed
during this study are included in this article
(and based on the references therein).
All the numerical calculations have been
carried out on the basis of a Fortran 90 compiler.

\bibliographystyle{mnras}

\begin{thebibliography}{99}

\bibitem[\protect\citeauthoryear{Ackermann}{2018}]{Ackermann2018}
Ackermann M., et al., 2018, ApJ, 857, 49

\bibitem[\protect\citeauthoryear{Barco}{2021}]{Barco2021}
Barco O., 2021, MNRAS, 506, 806

\bibitem[\protect\citeauthoryear{Barco}{2022}]{Barco2022}
Barco O., 2022, MNRAS, 512, 2925

\bibitem[\protect\citeauthoryear{Bartolo}{2019}]{Bartolo2019}
Bartolo N., De Luca V., Franciolini G., Peloso M., Racco D., Riotto A., 2019,
Phys. Rev. D, 99, 103521

\bibitem[\protect\citeauthoryear{Belotsky}{2019}]{Belotsky2019}
Belotsky K. M., et al., 2019, Eur. Phys. J. C, 79, 246

\bibitem[\protect\citeauthoryear{Boumaza}{2021}]{Boumaza2021}
Boumaza H., 2021, Eur. Phys. J. C, 81, 448

\bibitem[\protect\citeauthoryear{Capela}{2013}]{Capela2013}
Capela F., Pshirkov M., Tinyakov P., 2013, Phys. Rev. D, 87, 123524

\bibitem[\protect\citeauthoryear{Carr}{1974}]{Carr1974}
Carr B. J., Hawking S. W., 1974, MNRAS, 168, 399

\bibitem[\protect\citeauthoryear{Carr}{2020}]{Carr2020}
Carr B., Kühnel F., 2020, Annu. Rev. Nucl. Part. Sci., 70, 355

\bibitem[\protect\citeauthoryear{Carr}{2021}]{Carr2021}
Carr B., Kohri K., Sendouda Y., Yokoyama J., 2021, Rep. Prog. Phys., 84,
116902

\bibitem[\protect\citeauthoryear{Carr}{2022}]{Carr2022}
Carr B., Kühnel F., 2022, SciPosts Phys. Lect. Notes, p. 48

\bibitem[\protect\citeauthoryear{Chang}{2021}]{Chang2021}
Chang L. J., Necib L., 2021, MNRAS, 507, 4715

\bibitem[\protect\citeauthoryear{Conselice}{2006}]{Conselice2006}
Conselice C. J., 2006, MNRAS, 373, 1389

\bibitem[\protect\citeauthoryear{Conselice}{2016}]{Conselice2016}
Conselice C. J., Wilkinson A., Duncan K., Mortlock A., 2016, ApJ, 830, 83

\bibitem[\protect\citeauthoryear{Coogan}{2021}]{Coogan2021}
Coogan A., Morrison L., Profumo S., 2021, Phys. Rev. Lett., 126, 171101

\bibitem[\protect\citeauthoryear{Cordes}{2016}]{Cordes2016}
Cordes J. M., Wasserman I., 2016, MNRAS, 457, 232

\bibitem[\protect\citeauthoryear{Dasgupta}{2020}]{Dasgupta2020}
Dasgupta B., Laha R., Ray A., 2020, Phys. Rev. Lett., 125, 101101

\bibitem[\protect\citeauthoryear{deAngelis}{2018}]{deAngelis2018}
De Angelis A., et al., 2018, J. High Energy Astrophys., 19, 1

\bibitem[\protect\citeauthoryear{deLuca}{2021}]{deLuca2021}
De Luca V., Franciolini G., Riotto A., 2021, Phys. Rev. Lett., 126, 041303

\bibitem[\protect\citeauthoryear{Akiyama}{2019}]{Akiyama2019}
Event Horizon Telescope Collaboration et al., 2019, ApJ, 875, L1

\bibitem[\protect\citeauthoryear{Akiyama}{2022}]{Akiyama2022}
Event Horizon Telescope Collaboration et al., 2022, ApJ, 930, L12

\bibitem[\protect\citeauthoryear{Fantina}{2020}]{Fantina2020}
Fantina A. F., De Ridder S., Chamel N., Gulminelli F., 2020, A\&A, 633,
A149

\bibitem[\protect\citeauthoryear{Fynbo}{2001}]{Fynbo2001}
Fynbo J. U., et al., 2001, A\&A, 369, 373

\bibitem[\protect\citeauthoryear{Hartle}{2003}]{Hartle2003}
Hartle J. B., 2003, Gravity : an introduction to Einstein’s general relativity

\bibitem[\protect\citeauthoryear{Hawking}{1971}]{Hawking1971}
Hawking S. W., 1971, MNRAS, 152, 75

\bibitem[\protect\citeauthoryear{Hawking}{1974}]{Hawking1974}
Hawking S. W., 1974, Nature, 248, 30

\bibitem[\protect\citeauthoryear{Hawking}{1975}]{Hawking1975}
Hawking S. W., 1975, Commun. Math. Phys., 43, 199

\bibitem[\protect\citeauthoryear{Inomata}{2017}]{Inomata2017}
Inomata K., Kawasaki M., Mukaida K., Tada Y., Yanagida T. T., 2017,
Phys. Rev. D, 96, 043504

\bibitem[\protect\citeauthoryear{Jakobsson}{2004}]{Jakobsson2004}
Jakobsson P., Hjorth J., Fynbo J. P. U., Watson D., Pedersen K., Björnsson
G., Gorosabel J., 2004, ApJ, 617, L21

\bibitem[\protect\citeauthoryear{Kohri}{2021}]{Kohri2021}
Kohri K., Terada T., 2021, Phys. Lett. B, 813, 136040

\bibitem[\protect\citeauthoryear{Kouvaris}{2008}]{Kouvaris2008}
Kouvaris C., 2008, Phys. Rev. D, 77, 023006

\bibitem[\protect\citeauthoryear{Laha}{2019}]{Laha2019}
Laha R., 2019, Phys. Rev. Lett., 123, 251101

\bibitem[\protect\citeauthoryear{Laha}{2020}]{Laha2020}
Laha R., Muñoz J. B., Slatyer T. R., 2020, Phys. Rev. D, 101, 123514

\bibitem[\protect\citeauthoryear{Lauer}{2021}]{Lauer2021}
Lauer T. R., et al., 2021, ApJ, 906, 77

\bibitem[\protect\citeauthoryear{McMaken}{2022}]{McMaken2022}
McMaken T., 2022, MNRAS, 511, 1218

\bibitem[\protect\citeauthoryear{Melandri}{2012}]{Melandri2012}
Melandri A., et al., 2012, MNRAS, 421, 1265

\bibitem[\protect\citeauthoryear{Miller}{2022}]{Miller2022}
Miller A. L., Aggarwal N., Clesse S., De Lillo F., 2022, Phys. Rev. D, 105,
062008

\bibitem[\protect\citeauthoryear{Montero-Camacho}{2019}]{Montero-Camacho2019}
Montero-Camacho P., Fang X., Vasquez G., Silva M., Hirata C. M., 2019,
J. Cosmology Astropart. Phys., 2019, 031

\bibitem[\protect\citeauthoryear{Motahar}{2017}]{Motahar2017}
Motahar Z. A., Blázquez-Salcedo J. L., Kleihaus B., Kunz J., 2017, Phys.
Rev. D, 96, 064046

\bibitem[\protect\citeauthoryear{Murata}{2006}]{Murata2006}
Murata K., Soda J., 2006, Phys. Rev. D, 74, 044018

\bibitem[\protect\citeauthoryear{Pani}{2014}]{Pani2014}
Pani P., Loeb A., 2014, J. Cosmology Astropart. Phys., 2014, 026

\bibitem[\protect\citeauthoryear{Papanikolaou}{2021}]{Papanikolaou2021}
Papanikolaou T., Vennin V., Langlois D., 2021, J. Cosmology Astropart.
Phys., 2021, 053

\bibitem[\protect\citeauthoryear{Pearson}{2018}]{Pearson2018}
Pearson J. M., Chamel N., Potekhin A. Y., Fantina A. F., Ducoin C., Dutta
A. K., Goriely S., 2018, MNRAS, 481, 2994

\bibitem[\protect\citeauthoryear{Pearson}{2019}]{Pearson2019}
Pearson J. M., Chamel N., Potekhin A. Y., Fantina A. F., Ducoin C., Dutta
A. K., Goriely S., 2019, MNRAS, 486, 768

\bibitem[\protect\citeauthoryear{Ray}{2021}]{Ray2021}
Ray A., Laha R., Muñoz J. B., Caputo R., 2021, Phys. Rev. D, 104, 023516

\bibitem[\protect\citeauthoryear{Read}{2014}]{Read2014}
Read J. I., 2014, J. Phys. G: Nucl. Part. Phys., 41, 063101

\bibitem[\protect\citeauthoryear{Ryde}{1999}]{Ryde1999}
Ryde F., 1999, Astrophys. Lett. Commun., 39, 281

\bibitem[\protect\citeauthoryear{Ryde}{2004}]{Ryde2004}
Ryde F., 2004, ApJ, 614, 827

\bibitem[\protect\citeauthoryear{Saha}{2022}]{Saha2022}
Saha A. K., Laha R., 2022, Phys. Rev. D, 105, 103026

\bibitem[\protect\citeauthoryear{Sartore}{2010}]{Sartore2010}
Sartore N., Ripamonti E., Treves A., Turolla R., 2010, A\&A, 510, A23

\bibitem[\protect\citeauthoryear{Smyth}{2020}]{Smyth2020}
Smyth N., Profumo S., English S., Jeltema T., McKinnon K., Guhathakurta
P., 2020, Phys. Rev. D, 101, 063005

\bibitem[\protect\citeauthoryear{Sofue}{2020}]{Sofue2020}
Sofue Y., 2020, Galaxies, 8, 37

\bibitem[\protect\citeauthoryear{Wang}{2021}]{Wang2021}
Wang S., Xia D.-M., Zhang X., Zhou S., Chang Z., 2021, Phys. Rev. D, 103,
043010

\bibitem[\protect\citeauthoryear{Yazadjiev}{2016}]{Yazadjiev2016}
Yazadjiev S. S., Doneva D. D., Popchev D., 2016, Phys. Rev. D, 93, 084038

\bibitem[\protect\citeauthoryear{Zeldovich}{1967}]{Zeldovich1967}
Zeldovich Y. B., Novikov I. D., 1967, Soviet Ast., 10, 602

\end{thebibliography}

\bsp
\label{lastpage}
\end{document}